\documentclass[aps,prb,a4paper, amsfonts, amssymb, amsmath, reprint, showkeys, nofootinbib, twoside]{revtex4-1}
\usepackage[english]{babel}
\usepackage[utf8]{inputenc}
\usepackage[colorinlistoftodos, color=green!40, prependcaption]{todonotes}
\usepackage{amsthm}
\usepackage{mathtools}
\usepackage{physics}
\usepackage{xcolor}
\usepackage{graphicx}
\usepackage[left=23mm,right=13mm,top=35mm,columnsep=15pt]{geometry} 
\usepackage{adjustbox}
\usepackage{placeins}
\usepackage[T1]{fontenc}
\usepackage{lipsum}
\usepackage{csquotes}
\usepackage{comment}
\usepackage[pdftex, pdftitle={Article}, pdfauthor={Author}]{hyperref} 
\usepackage{ulem}
\usepackage{xcolor}
\DeclareRobustCommand{\erase}{\bgroup\markoverwith{\textcolor{red}{\rule[.5ex]{2pt}{0.4pt}}}\ULon}

\begin{document}
\title{X-ray magnetic circular dichroism originating from the $T_z$ term in collinear altermagnets under trigonal crystal field}

\author{Norimasa Sasabe$^{1}$}
\email[Correspondence email address: ]{SASABE.Norimasa@nims.go.jp}
\author{Yuta Ishii$^{1}$}
\author{Yuichi Yamasaki$^{1,2,3}$}
\email[Correspondence email address: ]{YAMASAKI.Yuichi@nims.go.jp}

\affiliation{$^{1}$Center for Basic Research on Materials, National Institute for Materials Science (NIMS), Tsukuba, 305-0047, Japan \\
$^{2}$International Center for Synchrotron Radiation Innovation Smart, Tohoku University, Sendai 980-8577, Japan \\
$^{3}$Center for Emergent Matter Science (CEMS), RIKEN, Wako 351-0198, Japan}

\date{\today}

\begin{abstract}
We investigate the microscopic origin and spectral features of X-ray magnetic circular dichroism (XMCD) in collinear antiferromagnets with trigonal crystal fields, using $\alpha$-MnTe as a prototypical example. 
Although such systems exhibit zero net magnetization, we demonstrate that XMCD can emerge from the anisotropic magnetic dipole operator $T_z$, arising from quadrupolar spin distributions. 
By constructing a complete multipole basis and analyzing the symmetry conditions under trigonal distortion, we identify specific spin and orbital configurations that enable a finite XMCD response. 
Further, we employ both one-electron and multi-electron models, including spin-orbit coupling and Coulomb interactions, to calculate the XMCD spectra for various $d^n$ configurations. 
Our findings provide theoretical benchmarks for XMCD in altermagnets and highlight the key role of orbital symmetry and magnetic anisotropy in realizing observable dichroic effects.
\end{abstract}

\keywords{altermagnet, x-ray magnetic circular dichroism, multipole, 3$d$ transition metal compounds}

\maketitle

\section{Introduction}
Antiferromagnetic materials with zero net magnetization have recently attracted significant attention as promising platforms for spintronic applications due to their ultrafast spin dynamics and insensitivity to external magnetic fields \cite{Nakatsuji2015, Jungwirth2016, Smejkal2018, Nakatsuji2022AnnualReview, Smejkal2022_PRX, Smejkal2022,  Smejkal2023_NatRevMater, DalDin2024}. However, the absence of macroscopic magnetic moments makes it inherently difficult to probe their microscopic spin arrangements.
Recent studies have revealed that even antiferromagnets can exhibit macroscopic phenomena typically associated with ferromagnets, such as the anomalous Hall effect \cite{Noda2016_PhysChemChemPhys, Mn3Sn_2017_Ikhlas_LargeNernst,  Okugawa_AM2018, Ahn2019PRB, Naka2019NatCom, Smejkal2020_SciAdv} and the magneto-optical Kerr effect \cite{Mn3Sn_2018_MOKE_higo, WimmerPRB2019, NakaPRB2020, ZhouPRB2021, YangPRB2021}, when they share the same magnetic point group symmetry as ferromagnets \cite{PhysRevB.95.094406, SuzukiPRB2019, Yatsushiro2021PRB}. 
These systems, often classified as alternating magnets, exhibit spin splitting and Berry curvature-induced responses, similar to those in ferromagnets \cite{Mn3Sn_2017_Wyle, HayamiJPSJ2019, hayami2021essential,AttiasPRB2024, HayamiJLPEA2024, OhgataPRB2025}.
Altermagnets possess magnetic-multipole symmetries determined by the point group of their antiferromagnetic order; the $s$-wave type corresponds to a magnetic dipole and the $d$-wave type to a magnetic toroidal quadrupole \cite{Koizumi2023, Koizumi_2025}, providing a basis for understanding their physical properties.

X-ray magnetic circular dichroism (XMCD), which measures the absorption difference between left- and right-circularly polarized X-rays, serves as a useful technique to probe magnetic moments selectively for each element and to estimate magnetic characteristics based on sum rules\cite{ Thole1992, Carra1993, carra1993magnetic,  TanakaJo1996JPSJ, Stohr1999, vanDerLaan2014}.
It has traditionally been regarded as a technique applicable only to ferromagnets and ferrimagnets \cite{XMCD_first, Schuetz1987, ChenPRB1990, Chen1995, ishii2024microscopic, YamasakiSTAMM2025}, while it has been demonstrated that certain antiferromagnets \cite{XMCD_Kimata, sakamoto2021observation, sakamoto2024bulk, 
LaanPRB2021, Hariki2024PRL, MnTe2024Nature, OkamotoAdvMat2024}, where the spin and orbital magnetic moments cancel out, the anisotropic magnetic dipole term, commonly referred to as the $T_z$ term \cite{Thole1992, Carra1993, Oguchi}, can survive and yield a distinct dichroic response \cite{Yamasaki2020, sasabe2023ferroic, YamasakiSTAM2025}.
Prominent examples include the chiral antiferromagnet MnSn$_3$ and the rutile-type collinear antiferromagnet RuO$_2$. 
In both cases, the magnetic point groups are identical to those of ferromagnets. 
Theoretical and experimental studies have consistently shown that the observed XMCD signals originate not from a net magnetic moment but from the anisotropic magnetic dipole term \cite{Yamasaki2020, Sasabe_PRL_Mn3Sn, XMCD_Kimata, sakamoto2021observation, sasabe2023ferroic, sakamoto2024bulk, KuritaJPSJ2024}.
This behavior can be understood in terms of orbital splitting induced by low-symmetry crystal fields, which results in a nonvanishing expectation value of the anisotropic magnetic dipole operator.
More recently, XMCD signals have also been observed and theoretically supported in the collinear antiferromagnet $\alpha$-MnTe \cite{Hariki2024PRL, MnTe2024Nature}, which crystallizes in the NiAs-type structure and possesses a crystal field with threefold rotational symmetry  \cite{GreenwaldMnTe1953, KomatsubaraJPSJ1963, kunitomi1964}. 
However, due to this symmetry, the role of the anisotropic magnetic dipole contribution remains ambiguous.
In this study, we aim to elucidate the relationship between the electronic structure and the anisotropic magnetic dipole contributions under a three-fold symmetric crystal field, taking antiferromagnetic ordering in the NiAs-type structure as a prototypical example.

\section{Complete multipole basis}
It is known that the electronic states associated with magnetic order can generally be expanded and described using a complete set of magnetic multipole basis functions \cite{kusunose2020complete, hayami2024unified, Hayami2024JPSJ}. 
On this basis, magnetic dipole moments are composed of spinless and spinful dipoles. The spinless multipoles can be described solely in terms of orbital angular momentum operators and thus correspond to orbital magnetic moments. 
On the other hand, spinful multipoles are obtained by coupling spinless multipoles with spin operators.
In general, multipole operators within the complete magnetic multipole basis can be expressed as:
\begin{equation}
   \hat{M}_{lm}^{(s,k)} =
     i^{s \pm k} \sum_n C^{lm}_{l\pm k, m-n; sn} \hat{X}_{l\pm k,m-n}^{\mathrm{(orb)}} \, \sigma_{sn}   
\end{equation}
Here, $s = 0$ corresponds to spinless multipoles, and $s = 1$ to spinful multipoles. $C_{l_1m_1;l_2m_2}^{l_3m_3}$ denotes the Clebsch-Gordan coefficients, and $\sigma_{sn}$ represents the Pauli matrices. 
The index $k$ indicates the allowed values for which the Clebsch-Gordan coefficient does not vanish: for $s = 0$, only $k = 0$ is relevant; for $s = 1$, only $k = -1, 0, 1$ yield non-zero contributions.
The magnetic dipole is even under spatial inversion and odd under time-reversal symmetry. 
To satisfy these symmetry requirements, the operator $X$ corresponds to charge a multipole ($Q_{lm}$) for $s = 1$, $k = \pm 1$, and to a electric toroidal multipole ($G_{lm}$) for $s = 1$, $k = 0$.
XMCD is sensitive to several components of the magnetic dipole: the spinless orbital magnetic moment, the spinful monopolar spin component formed by coupling a charge monopole with spin, and the quadrupolar spin component formed by coupling a quadrupole with spin. 
In particular, by applying sum rules, it is possible to separate and quantify the spinless and spinful multipole contributions \cite{YamasakiSTAM2025}.

\begin{figure}[t]
    \centering
    \includegraphics[width=0.48\textwidth]{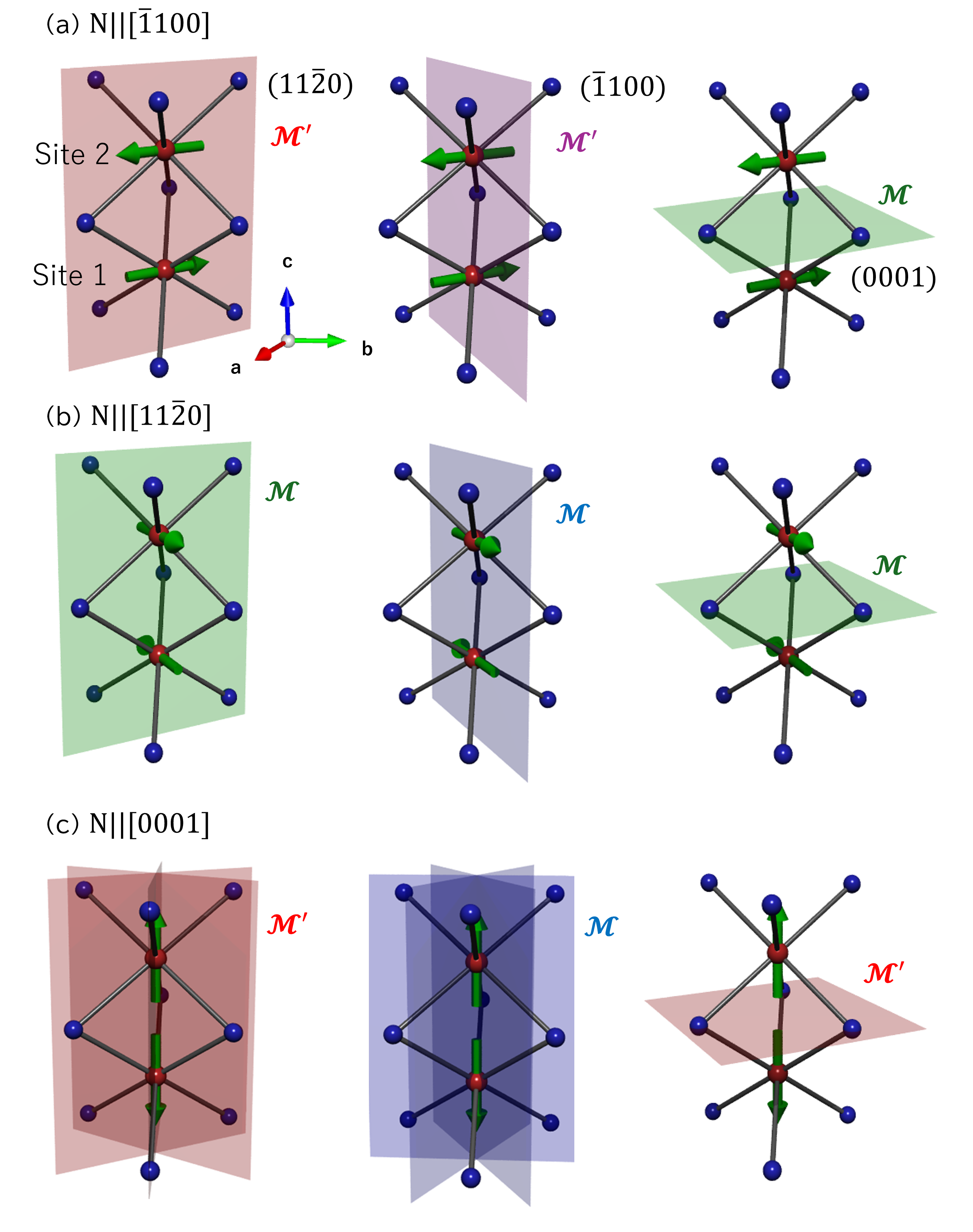}
    \caption{
    Lattice of the NiAs-type structure, which is the same as $\alpha$-MnTe, and symmetry operations in a collinear antiferromagnetic state. 
    Each panel illustrates a specific symmetry operation acting on the magnetic structure, where green arrows represent antiparallel magnetic moments with the Neel vector parallel to (a) the $[\bar{1}100]$, (b) the $[11\bar{2}0]$ and (c) the [0001] direction. 
    Green and red planes indicate mirror ($\mathcal{M}$) and mirror with the time reversal ($\mathcal{M}^\prime$), respectively.
    Blue and purple planes for the $\langle 1\bar{1}00\rangle$ denote the $c$-glide mirror planes.}
    \label{fig:magneticstructure}
\end{figure}

The quadrupolar spin operator, often referred to as the anisotropic magnetic dipole operator or $t_z$ operator, plays a central role in describing anisotropic spin distributions that contribute to the XMCD signal even in systems with vanishing net magnetization. This operator captures the coupling between the spin degree of freedom and anisotropic charge distributions characterized by quadrupole moments, and is particularly relevant in antiferromagnetic or orbitally degenerate systems where conventional spin and orbital magnetic moments cancel out macroscopically.
Formally, the rank-1 dipole operator is defined as the tensor product between a rank-2 quadrupole tensor $\mathbf{Q}$ and the spin vector $\mathbf{s}$: $\mathbf{T} = \mathbf{Q} \cdot \mathbf{s}$.
Here, $\mathbf{Q}$ is a traceless symmetric tensor constructed from the position vector $\mathbf{r}$ as
\begin{equation}
Q_{\alpha\beta} = \delta_{\alpha\beta} - 3 \hat{r}_\alpha \hat{r}_\beta,
\end{equation}
where $\hat{\mathbf{r}} = \mathbf{r}/|\mathbf{r}|$ denotes the unit vector in real space, and $\delta_{\alpha\beta}$ is the Kronecker delta. This form reflects the characteristic spatial anisotropy of quadrupolar distributions.
It must be emphasized that the quadrupolar spin operator is in fact a dipole operator, and thus should be clearly distinguished from magnetic quadrupole and magnetic octupole moments.

To facilitate multiplet analysis and coupling with angular momentum states, it is convenient to recast the operator in terms of spherical tensor components. 
Using a set of normalized spherical harmonics $c_m^{(2)}$ for $m = -2, -1, 0, 1, 2$, the operator components $t_\pm$ and $t_z$ are given by \cite{Oguchi}:
\begin{align}\label{toperator}
t_\pm &= c_0^{(2)} s_\pm - \sqrt{6} \, c_{\pm 2}^{(2)} s_\mp \pm \sqrt{6} \, c_{\pm 1}^{(2)} s_z, \\
t_z &= -\sqrt{\frac{3}{2}} \, c_{-1}^{(2)} s_+ + \sqrt{\frac{3}{2}} \, c_1^{(2)} s_- - 2 c_0^{(2)} s_z,
\end{align}
where the spin operators are defined by $s_\pm = s_x \pm i s_y$ and $s_z$.
This decomposition makes explicit the selection rules and symmetry properties of each component in the context of dipole transitions. The presence of these terms enables finite XMCD signals even in collinear antiferromagnets, as they represent symmetry-allowed spin-orbital configurations that survive the cancellation of net moments.

\begin{figure}[t]\label{fig_trigonal}
    \centering
    \includegraphics[width=0.48\textwidth]{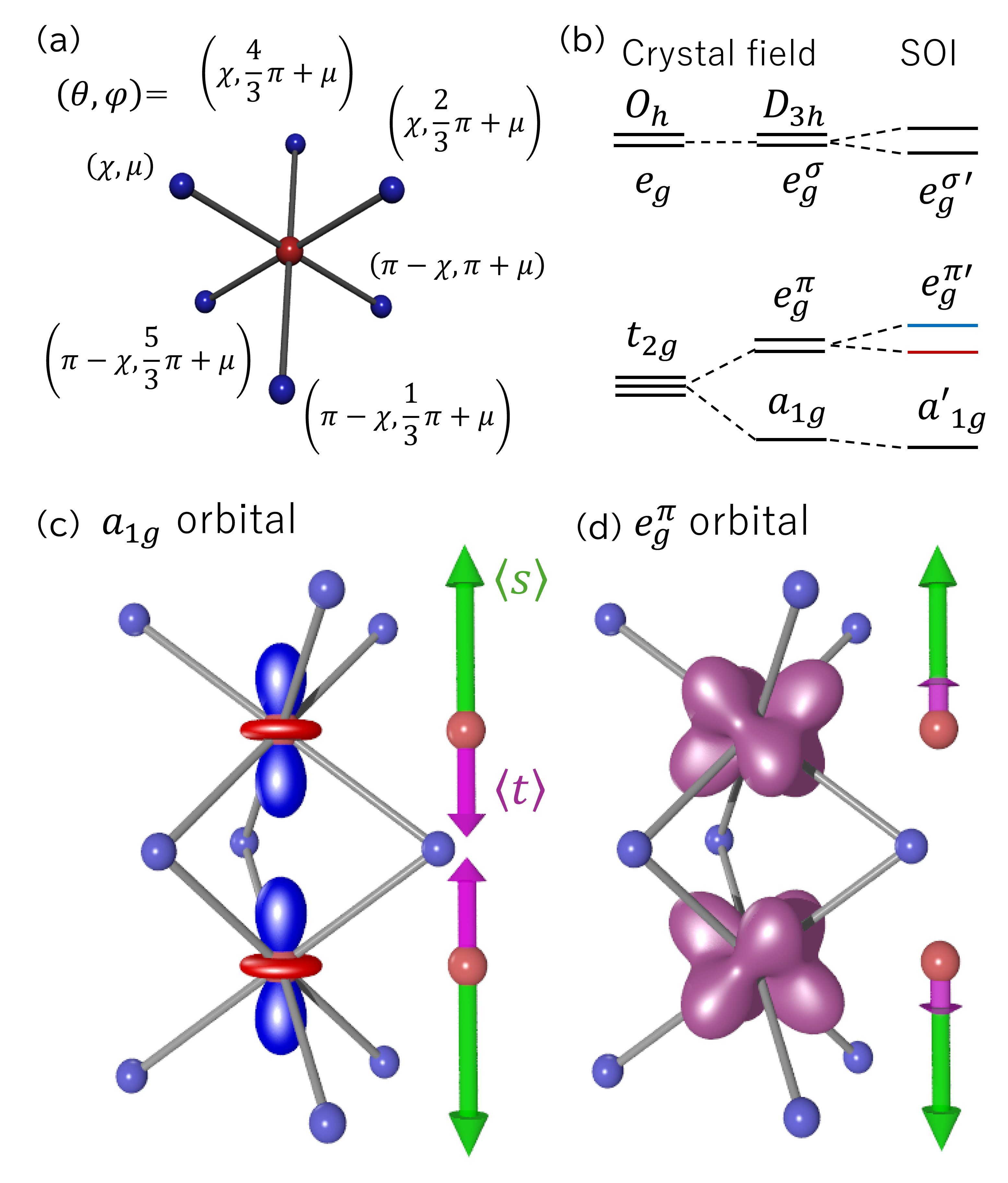}
    \caption{ Crystal field splitting of $d$-orbitals under ideal octahedral ($O_h$) symmetry and its modification in a trigonal crystal field with $D_{3d}$ symmetry. 
    (a) a schematic of atomic positions in a trigonally distorted octahedron, parameterized by spherical coordinates $(\theta, \varphi)$.
    Here, $\chi$ and $\mu$ describe the angular deviation from the ideal octahedral geometry, introducing threefold rotational symmetry around the central axis.
    (b) The energy level diagram: in $O_h$, the $d$-orbitals split into $e_g$ and $t_{2g}$, and under trigonal distortion, these further split into $e_g^\sigma$, $e_g^\pi$, and $a_{1g}$ states.
    Orbital shapes of (c) $a_{1g}$ and (d) $e_{g}^{\pi}$, and expectation value vectors of spin (green) and anisotropic magnetic dipole (purple) in antiferromagnetic structure with the N\'eel vector along the [0001] axis.
    }
    \label{fig:trigonal}
\end{figure}

\begin{figure*}[t]
    \centering
    \includegraphics[width=0.95\textwidth]{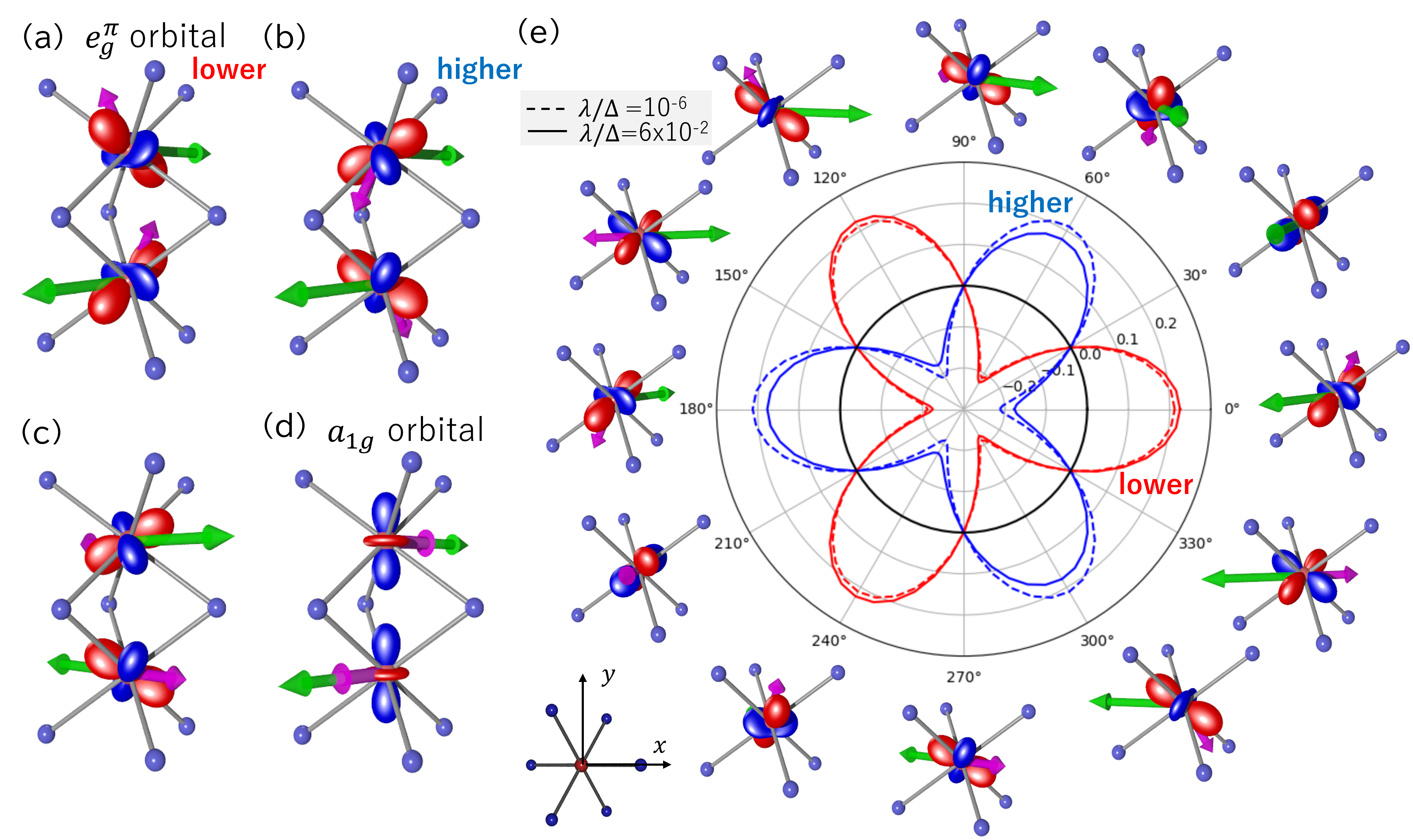}
    \caption{Schematic illustrations of the $\textbf{t}$ vector and orbital shapes for (a) the lower-energy, (b) the higher-energy $e_g^\pi$ states with $N||[\bar{1}100]$, (c) the lower $e_g^\pi$ state with $N||[0010]$ and (d) $a_{1g}$ state.  
    The green arrows indicate the spin direction, and the purple arrows represent the direction of the $\mathbf{t}$ vector.  
    (e) Angular dependence of the $t_z$ component with the N\'eel vector aligning within the basal (0001) plane, visualized as a polar plot.  
    The red and blue curves correspond to the expectation value of $\langle t_z\rangle$ in the lower- and higher-energy $e_g^\pi$ states, respectively.
    Surrounding structures illustrate how the orbital shape and $\mathbf{t}$ vector evolve with the in-plane spin direction at site 1.  
    The residual $t_z$ component survives when the N\'eel vector $\mathbf{N}$ is along $\langle \overline{1}100 \rangle$-type directions, whereas it vanishes along $\langle 1000 \rangle$-type directions.
    The dashed and solid lines represent the cases of $\lambda/\Delta = 10^{-6}$ and $\lambda/\Delta = 6\times 10^{-2}$, respectively.}
\label{fig:tz_vectore}
\end{figure*}

\section{Collinear Antiferromagnet in the NiAs-type structure}
In this paper, we investigate the XMCD response in the antiferromagnetic compound $\alpha$-MnTe
which crystallizes in the NiAs-type structure, belonging to the hexagonal crystal system with space group $P6_3/mmc$ (No. 194) [see Figure \ref{fig:magneticstructure}]. 
In this structure, Mn atoms occupy the $2a$ Wyckoff positions at $(0, 0, 0)$, while Te atoms are located at the $2c$ positions at $(\frac{1}{3}, \frac{2}{3}, \frac{1}{4})$. 
The Mn atoms form stacked hexagonal layers, with each Mn atom octahedrally coordinated by six Te atoms. 
This leads to the formation of linear chains of Mn atoms running along the crystallographic $c$-axis, separated by Te atoms.
This arrangement supports collinear antiferromagnetic ordering, in which the magnetic moments of adjacent Mn layers alternate in direction. 
The interplay between hexagonal symmetry and antiferromagnetic ordering breaks certain spatial and time-reversal symmetries, resulting in a magnetic space group that induces the spin splitting in the $k$-space.

In the collinear antiferromagnetic phase of a material with the same crystal structure as $\alpha$-MnTe, certain symmetry operations are either preserved or broken due to the magnetic ordering. 
Figure \ref{fig:magneticstructure} summarizes these operations using visual representations. 
The green arrows correspond to the antiparallel spin orientations on the magnetic ions, while the overlaid planes indicate symmetry elements. 
Specifically, red planes denote mirror operations that require time-reversal symmetry to be maintained ($\mathcal{M}^\prime$), while green planes correspond to ordinary mirror operations ($\mathcal{M}$). 
Noted that the $(\bar{1}100)$ plane, along with its crystallographically equivalent counterparts, indicated in blue and purple, constitutes a set of $c$-glide mirror planes and that with the time-reversal symmetry.

When the N\'eel vector, defined as the difference between the two antiferromagnetic spin sublattices, is oriented along the $[\bar{1}100]$ direction ($N||[\bar{1}100]$) as shown in Fig. \ref{fig:magneticstructure}(a)], there exists one mirror plane $\mathcal{M}$ and two perpendicular mirror planes $\mathcal{M}^\prime$. 
This configuration preserves the same magnetic point group symmetry as a magnetic dipole, thereby classifying the system as an altermagnet \cite{LoveseyPRB2023, AoyamaPRM2024}. 
This magnetic structure corresponds to the condition under which XMCD signals have been experimentally observed, and its pseudo-magnetic dipole moment aligns along the [0001] direction.
In contrast, when $\mathbf{N} \parallel [11\bar{2}0]$, the presence of three equivalent mirror planes $\mathcal{M}$ leads to a conventional antiferromagnetic structure. 
In this case, the symmetry does not support XMCD activity due to the absence of magnetic dipole-like symmetry breaking.
When $\mathbf{N} \parallel [0001]$, there again exists one mirror plane $\mathcal{M}$ and two orthogonal mirrors $\mathcal{M}^\prime$, indicating alternating magnet character. 
However, due to the presence of threefold rotational symmetry about the $[0001]$ axis, the net magnetic dipole moment vanishes by symmetry. 
As a result, XMCD is not allowed in this configuration \cite{LiPRB2025}. 
Nevertheless, this state belongs to the so-called $g$-type altermagnet, which causes a 
momentum-dependent spin splitting \cite{Smejkal2023_NatRevMater, DingPRL2024, YangNatCom2025, GalindezAdM2025}.

\section{Collinear antiferromagnet under trigonal crystal field}
We investigate the microscopic origin of the XMCD signal expected from the magnetic symmetry of the collinear antiferromagnet with the NiAs-type crystal structure, where two octahedra share their face, with particular focus on the expectation value of the quadrupolar spin component.
The atomic positions surrounding the magnetic ion can be represented in polar coordinates $(\theta, \varphi)$ expressed in terms of two angles, $\chi$ and $\mu$, as shown in Fig.\ref{fig:trigonal}(a), with the [0001] direction taken as the $z$-axis and the $[1\bar{1}00]$ direction as the $x$-axis.
Site 1 (2) corresponds to $\mu = 0$ ($\mu = \pi$).
In an ideal octahedral environment with $O_h$ symmetry, where the angle $\chi$ between the ligand directions and the coordinate axes satisfies $\chi = \arccos(1/\sqrt{3})$, the five degenerate $d$-orbitals of a transition metal ion split into two sets: a doubly degenerate $e_g$ level and a triply degenerate $t_{2g}$ level. 
When a trigonal distortion is introduced along the threefold axis, reducing the symmetry to $D_{3h}$, these orbitals further split into non-degenerate or differently degenerate states: the $e_g$ level becomes $e_g^\sigma$, while the $t_{2g}$ level splits into $a_{1g}$ and $e_g^\pi$ components.
When $\chi > \arccos\left( {1}/{\sqrt{3}} \right)$, as in the case of $\alpha$-MnTe, the $a_{1g}$ level becomes lower in energy than the $e_g^\pi$ level [see Fig. \ref{fig:trigonal}(b)].

When the spin is aligned along the [0001] direction, the crystal possesses threefold symmetry.  
Expressing the $d$-orbital wavefunctions using spherical harmonics $Y_{m}^{(l)}(\hat{\textbf{r}})$ ($l=2$) and the radial function $R_d(r)$ as $\lvert m \rangle = Y_{m}^{(2)}(\hat{\textbf{r}}) R_d(r)$, the $a_{1g}$ state corresponds to $\lvert 0 \rangle$, and the $e_g^\pi$ states are represented by
\begin{equation}
    |\pi_\pm \rangle = \cos\zeta|\pm 2\rangle \mp \eta_j \sin \zeta|\mp 1\rangle
\end{equation}
with $\eta$ and $\zeta$ depending on the trigonal distortion.
The parameter $\eta$ distinguishes between the two inequivalent sites, taking the value $\eta = +1$ at site 1 and $\eta = -1$ at site 2.  
Under ideal octahedral ($O_h$) symmetry, the associated physical quantity satisfies the condition $\tan \zeta = 1/\sqrt{2}$.
Figure 2(c) and 2(d) illustrate the spatial distribution of each orbital state. 
The $a_{1g}$ orbital, being real-valued, is visualized with red and blue colors to represent its phase.  
On the other hand, the $e_g^\pi$ orbitals are complex-valued, and thus their visualizations show the magnitude of the wavefunction.
It should be noted here that the $e_g^\pi$ orbitals retain residual orbital angular momentum, $\langle \pi_\pm |\hat{l}_z|\pi_\pm\rangle = \pm (3\cos^2\zeta-1)$, as it is not fully quenched.
The expectation value of the operator $t_z$ in Eq. (\ref{toperator}) for the $a_{1g}$ orbital state is obtained as $\langle a_{1g};\sigma_z =\pm 1/2|t_z|a_{1g};\pm 1/2 \rangle = \mp 2/7$.
On the other hand, in the basis $\{\pi_+;1/2,\, \pi_+;-1/2,\, \pi_-;1/2,\, \pi_-;-1/2\}$, the expectation value of $t_z$ is represented as
\begin{equation}
\langle t_z \rangle=
\frac{1}{7} 
\begin{pmatrix}
\tau(\zeta) & 0 & 0 & 0 \\
0 & -\tau(\zeta) & 3 \eta_j\sin2\zeta & 0 \\
0 & 3 \eta_j\sin2\zeta & \tau(\zeta) & 0 \\
0 & 0 & 0 & -\tau(\zeta)
\end{pmatrix}    
\end{equation}
with $\tau (\zeta) \equiv 3\cos^2\zeta - 1$.
These results indicate that when $\mathbf{N}||[0001]$, the $t_z$ term of $a_{1g}$ orbital is oriented antiparallel to the spin, whereas the $e_g^\pi$ orbitals are parallel to it as shown in Fig. \ref{fig:trigonal}(c) and (d)]. 
Therefore, similar to the spin moment, the $t_z$ contribution is effectively canceled in the antiferromagnetic order.  
Consequently, in a $g$-wave altermagnet XMCD is not expected to emerge due to the cancellation of both spin and $t_z$ contributions.
On the other hand, due to the presence of off-diagonal components in the $t_z$ operator, a nonzero expectation value of XMCD may emerge when the spin is oriented within the (0001) plane.

\begin{table}[t]
    \begin{tabular}{r r r r c c}
\hline\hline
$d^{n}$\,\, & $L_{\rm z}$\,\,\,\, & $S_{\rm z}$\,\,\,\, & $T_{\rm z}$\,\,\,\,& $\lambda_X$ (meV) & $\Delta E$ (meV)  \\
\hline\hline
Ti$^{2+}$ ($3d^2$) & -0.0008 & 0.0000 & 0.0051  &  16 & 65  \\
V$^{2+}$  ($3d^3$) &  0.0000 & 0.0000 & 0.0000  & 27 & 600  \\
Cr$^{2+}$ ($3d^4$) & -0.0378 & 0.0018 & -0.2211 & 30 & 0.5 \\
Mn$^{2+}$ ($3d^5$) &  0.0000 & 0.0000 & 0.0000  & 40 & 600  \\
Fe$^{2+}$ ($3d^6$) & -0.0098 & -0.0006 & 0.0235  & 52 & 80  \\
Co$^{2+}$ ($3d^7$) &  0.0268 & 0.0016 & -0.0172 & 66 & 106  \\
Ni$^{2+}$ ($3d^8$) & -0.0001 & 0.0000 & -0.0002 & 74 & 597  \\
Cu$^{2+}$ ($3d^9$) &  0.0896 & 0.0023 & 0.1910  & 102 & 0.7  \\
\hline\hline
\end{tabular}
\caption{Expectation values of orbital angular momentum $L_z$, spin angular momentum $S_z$, and anisotropic magnetic dipole operator $T_z$, along with the spin-orbit coupling constant ($\lambda_X$) and the energy difference $\Delta E$ between the ground and the first excited states, for divalent $3d$ transition-metal ions ($X=$Ti$^{2+}$ to {Cu}$^{2+}$) in antiferromagnetic configurations under the trigonal state. 
Despite vanishing net magnetization, finite XMCD signals can appear due to nonzero $T_z$, particularly in configurations where orbital degeneracy is lifted by Coulomb interactions and the spin-orbit interaction.
}
\end{table}

\begin{figure*}[t]
    \centering
    \includegraphics[width=0.98\textwidth]{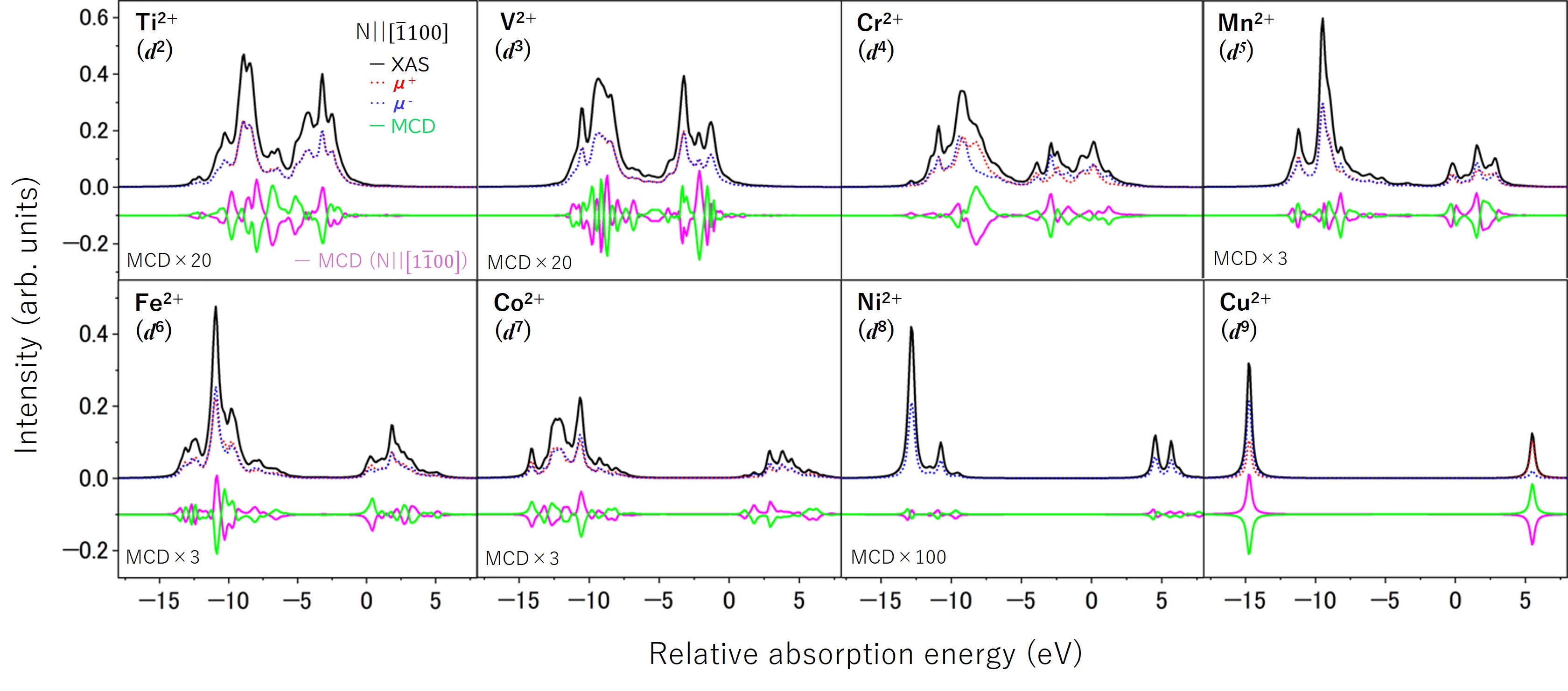}
    \caption{Calculated X-ray absorption spectra (XAS, black curves) and X-ray magnetic circular dichroism (XMCD, green curves) for divalent $3d$ transition-metal ions ($X=\mathrm{Ti}^{2+}$ to $\mathrm{Cu}^{2+}$) in antiferromagnetic configurations. 
    The dotted red and blue lines correspond to left- and right-circularly polarized absorption, respectively. 
    XMCD signals are scaled by factors indicated in each panel to enhance visibility. 
    All spectra are plotted relative energy to the absorption edge of each ion.
    }
    \label{fig:xas_xmcd}
\end{figure*}

To clarify the electronic structure when the N\'eel vector aligns within the (0001) basal plane, we consider an effective one-electron Hamiltonian.  
In the basis of $\lvert m\sigma \rangle$, we introduce the following simplified model:
\begin{equation}
    \mathcal{H}_{\text{eff}} = \sum_i V_\text{tri}^{(i)}+\lambda \textbf{l}_i \cdot \textbf{s}_i + g\textbf{s}_i\cdot\textbf{h}_\text{MF}^{(i)}
    \label{eq:one-electronHamiltonian}
\end{equation}
where $V_\text{tri}$ indicates the $i$-th site crystal field, $\lambda$ is the spin--orbit coupling constant, and $g$ is the Land\'{e} $g$-factor.
Here, $V_{\rm tri}$ is determined by considering the one-electron potential of $D_{3h}$ symmetry, expressed as
\begin{eqnarray}
V_{\rm tri}=& \sqrt{\frac{4 \pi}{5}}B^2_0Y^{(2)}_0+\sqrt{\frac{4 \pi}{9}}B^4_0Y^{(4)}_0 \notag \\
&+\sqrt{\frac{4 \pi}{9}}B^4_3\left(Y^{(4)}_{3}+Y^{(4)}_{-3}\right).
\end{eqnarray}
with $Y^{(l)}_m$ being the spherical harmonic function.
In this study, we adopt the crystal field parameters (in eV) as $B^2_0 = -0.56$, $B^4_0 = -1.92$, and $B^4_3 = \pm 2.88$.
The sign of $B^4_3$ determines the trigonal crystal field of site 1 or 2.
$\textbf{h}_{\mathrm{MF}}^{(i)}$ denotes the internal magnetic field within the mean-field approximation, and is assumed to be applied along $(-\cos\phi,-\sin\phi,0)$ within the (0001) plane.
We diagonalized the Hamiltonian and calculated the expectation value of the $\mathbf{t}$ vector for its ground state.  
Focusing on the $e_g^\pi$ orbitals, we found that in the absence of spin-orbit coupling, a twofold degeneracy remains.  
However, when the spin--orbit interaction is finite, this degeneracy is lifted, and the ground state takes a real-valued orbital state.
For the case of $\mathbf{N} \parallel [\overline{1}100]$, the lower energy $e_g^\pi$ orbital takes the form shown in Fig. \ref{fig:tz_vectore}(a), where the corresponding $\mathbf{t}$ vectors are illustrated in purple.  
The orbitals on neighboring sites are related by a 180-degree rotation, which results in a non-cancelled $t_z$ component.  
This residual $t_z$ is the origin of the finite XMCD signal when the incident x-ray is parallel to the [0001] axis.
It is also evident that the in-plane components of the $\mathbf{t}$ vectors cancel out between the two sites.  
Figure \ref{fig:tz_vectore}(b) additionally shows the other $e_g^\pi$ orbital, for which the $t_z$ component remains in the opposite direction, again without cancellation.  
In the absence of spin-orbit interaction, where the $e_g^\pi$ states are fully degenerate, the $t_z$ contributions cancel each other out within the $e_g^\pi$, and XMCD does not appear.

\begin{figure}[t]
    \centering
    \includegraphics[width=0.48\textwidth]{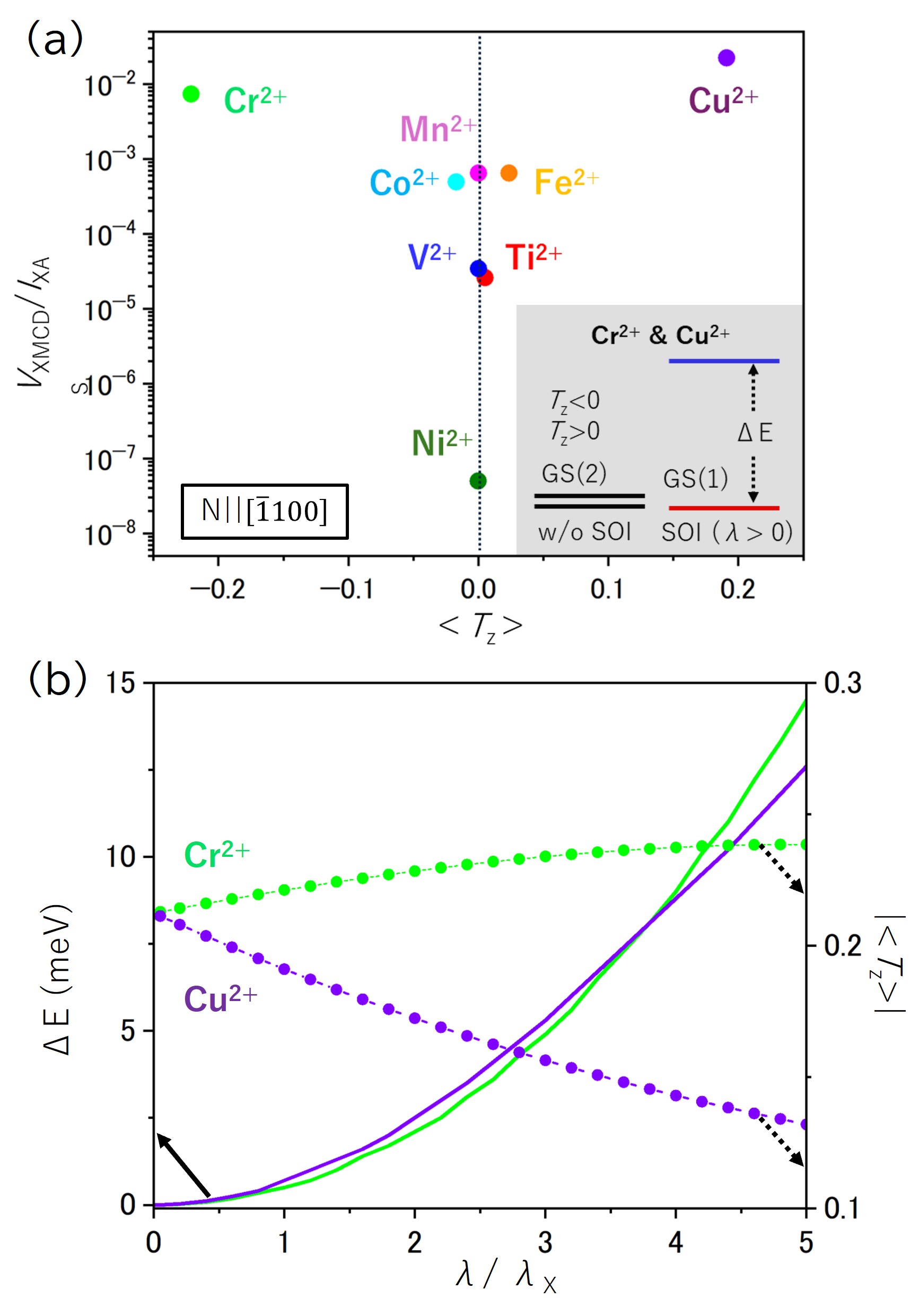}
    \caption{(a) Scatter plot comparing the expectation value of the magnetic dipole operator $T_z$ and the variance of the XMCD signal, $V_{\mathrm{XMCD}}$, defined as the energy-integrated square of the XMCD spectrum. It is normalized by the integrated XAS intensity $I_{\mathrm{XAS}}$. (b) For Cr$^{2+}$ and Cu$^{2+}$ ions, the dependence of the $T_z$ expectation value and the energy gap $\Delta E$ between the ground and first excited states on the spin-orbit coupling constant ($\lambda$) normalized by the calculated value on $X^{2+}$ ($\lambda_X$; $X=$ Cr and Cu).
}
    \label{fig:xas_xmcd_Cr1}
\end{figure}

\begin{figure}[t]
    \centering
    \includegraphics[width=0.48\textwidth]{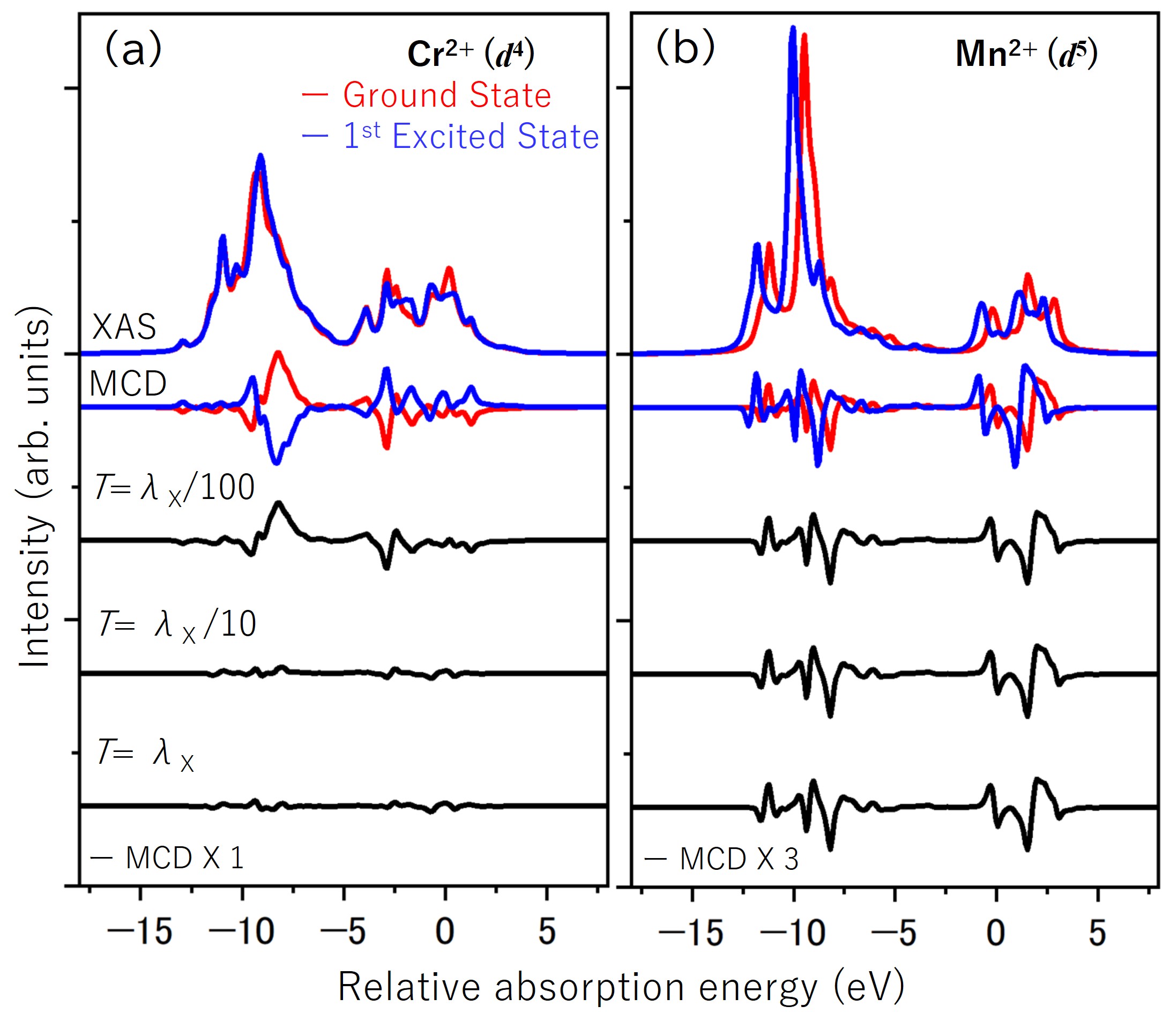}
    \caption{
    XMCD spectra of the excited states for (a) Cr$^{2+}$ ($3d^4$) and (b) Mn$^{2+}$ ($3d^5$). 
    Red and blue lines represent the XMCD signals calculated for the ground state and the first excited state, respectively. 
    The lower black curves show the XMCD spectra obtained by Boltzmann-weighted averaging over initial states at various temperatures ($T=\lambda_X/100,~\lambda_X/10, ~\lambda_X$; $X=$Cr$^{2+}$, Cu$^{2+}$). 
    The results indicate that the XMCD signal for Cr$^{2+}$ rapidly diminishes with increasing temperature due to a small energy gap $\Delta E$, whereas Mn$^{2+}$ shows a more robust response owing to a larger $\Delta E$ and a similar spectral shape for the two states.
    }
    \label{fig:xas_xmcd_Cr2}
\end{figure}

When $\mathbf{N} \parallel [11\bar{2}0]$, the electronic configuration of the lower energy $e_g^\pi$ state takes the form shown in Fig. \ref{fig:tz_vectore}(c). 
In this case, the $\mathbf{t}$ vector is antiparallel to the spin direction, resulting in no $t_z$ component and, consequently, the absence of XMCD.  
This indicates that a finite $t_z$ component remains when the N\'eel vector lies along directions equivalent to $\langle \overline{1}100 \rangle$,  
whereas it is canceled when aligned along directions equivalent to $\langle 1000 \rangle$. 
In addition, the $\mathbf{t}$ vector of the pure $a_{1g}$ state always points in the same direction as $\mathbf{s}$ as shown in Fig. \ref{fig:tz_vectore}(d), and therefore does not contribute to XMCD.
However, in reality, due to hybridization with the \( e_g \) states caused by the spin-orbit interaction, it does not vanish entirely.

Figure \ref{fig:tz_vectore}(e) summarizes the angular dependence of $t_z$ as a polar plot for different in-plane spin orientations.  
The red and blue lines represent the lower- and higher-energy $e_g^\pi$ states, respectively.  
It is evident that $t_z$ remains finite for spin directions along $\langle \overline{1}100 \rangle$, while it cancels out for directions along $\langle 1000 \rangle$.  
This confirms that although the stable shape of the orbital quadrupole depends on the spin direction, the $\mathbf{t}$ vector responds consistently with the symmetry of the system.
The dashed line corresponds to the limit of weak spin-orbit interaction ($\lambda/\Delta =10^{-6}$), in which the energy levels are degenerate and the oppositely oriented \( t_z \) components cancel each other out. 
In contrast, when the spin-orbit interaction is present ($\lambda/\Delta =6 \times 10^{-2}$), the degeneracy is lifted, the expectation values of \( t_z \) become different, and the effect becomes observable as XMCD.

\section{XMCD spectra calculation}
To explore the possibility of XMCD in more realistic materials beyond the one-electron approximation and to examine the XMCD spectra, we calculate the ground state and dipole transitions in a multi-electron system.
By extending Eq.~\ref{eq:one-electronHamiltonian} to a multi-electron system on the Fock basis of $d^n$ and introducing the Coulomb interaction, we obtain the initial state using the Lanczos method. 
Based on dipole transitions from this state, we calculate the XMCD spectrum within the atomic model, including full-multiplet effects.
The spin-orbit coupling constants and Slater integrals were obtained from ionic calculations based on the Hartree-Fock-Slater (HFS) method \cite{cowan1981theory}.
For the Slater integrals, 80 \% of the HFS values were used \cite{okada1989multiplet, tanaka1992temperature,  taguchi1997theory, matsubara2000polarization}. 
Consistent with Eq.~\ref{eq:one-electronHamiltonian}, the atomic model adopts the crystal-field parameters given in Eq.~(8) and an effective magnetic field of 0.3 eV arising from the antiferromagnetic interaction with $\mathbf{N} \parallel [\overline{1}100]$. 
We calculated the models for divalent 3$d$ transition-metal ions from Ti$^{2+}$ ($d^2$) to Cu$^{2+}$ ($d^9$), and the expectation values of $L_{\rm z}$, $S_{\rm z}$, and $T_{\rm z}$ in the initial ground state as summarized in Table 1.

Figure~\ref{fig:xas_xmcd} shows the calculated X-ray absorption spectra (XAS) with circular polarizations ($\mu_+$ and $\mu_-$) and XMCD for a series of $3d$ transition-metal ions in divalent states, assuming an antiferromagnetic configuration with $N||[\bar{1}100]$. 
Despite the collinear antiferromagnetic alignment, finite XMCD signals are observed along the [0001] direction in all cases, as shown in the green curves scaled for clarity.
The magenta line shows that the XMCD spectrum is completely reversed when the antiferromagnetic N\'eel vector is flipped as $N||[1\bar{1}00]$.
Since the XMCD spectrum allows the observation of each electronic state with energy resolution, it can be observed regardless of the number of electrons, even if the expectation value of $t_z$ operator integrated over the entire $d$ orbital is $\langle T_z\rangle=0$ such as in V$^{2+}$ and Mn$^{2+}$ ions (see Table 1).
In the final states excited by dipole transitions, the orbital degeneracy is lifted, leading to the conclusion that the XMCD in $\alpha$-MnTe originates from the $T_z$ term \cite{Hariki2024PRL, MnTe2024Nature}.
In an altermagnet FeS \cite{FeS2025NatMat}, which contains Fe$^{2+}$ ions under the same crystal field symmetry and shows a collinear antiferromagnetic order, the emergence of XMCD is similarly expected.

To clarify the relationship between the XMCD intensity and $T_z$ term, Fig. \ref{fig:xas_xmcd_Cr1}(a) displays the variance of the XMCD spectra plotted against the expectation value of $T_z$ in the ground state. 
To normalize the change in the number of holes, the XMCD variance is normalized by the integrated intensity of the XAS, and hereafter, we refer to this as the XMCD intensity. 
The XMCD intensity exhibits characteristic behavior reflecting each electronic state.
Based on the one electron approximation as discussed above, when either the $e_g$ or $t_{2g}$ orbitals in the $O_h$ crystal field are completely occupied (as in the cases of $d^3$, $d^5$, or $d^8$), the XMCD becomes weak, whereas it tends to become strong when they are partially occupied. 
In fact, the weakest XMCD signal is observed for $d^8$ Ni$^{2+}$, corresponding to a state with two holes in the nearly degenerate $e_g^\sigma$ orbitals. 
A similar situation might be expected for $d^3$ (V$^{2+}$), where the $t_{2g}$ orbitals are filled with three electrons. 
However, due to spin-orbit interaction, mixing with the minority-spin states occurs, resulting in partially occupied $t_{2g}$ orbitals and a stronger XMCD signal compared to Ni$^{2+}$. 
In the case of $d^5$ (Mn$^{2+}$) as well, since both the $t_{2g}$ and $e_g$ orbitals are partially occupied due to the spin-orbit interaction, the XMCD appears relatively strong.

In the case of $d^6$ (Fe$^{2+}$), where the $t_{2g}$ orbitals are partially occupied, a minority-spin electron enters the $a_{1g}$ orbital and two holes are present in the $e_g^{\pi}$ orbitals, leading to a weaker XMCD. 
However, due to spin-orbit interaction, mixing between $a_{1g}$ and $e_g^{\pi}$ occurs, resulting in the appearance of XMCD. 
In contrast, for $d^2$ (V$^{2+}$) and $d^7$ (Co$^{2+}$), one hole is located in the $e_g^{\pi}$ orbitals; therefore, a strong XMCD would be expected, but in reality it does not appear to be so strong. 
Examination of the ground states of the initial configurations reveals that a state in which holes predominantly occupy $a_{1g}$ is stabilized. 
This is because it is energetically more favorable, in terms of intra-atomic Coulomb interaction, for two electrons to occupy the same $e_g^{\pi}$ orbital rather than distributing one electron in $a_{1g}$ and one in $e_g^{\pi}$ orbitals.

The strongest XMCD is calculated for $d^4$ (Cr$^{2+}$) and $d^9$ (Cu$^{2+}$), where the $e_g^{\sigma}$ orbitals are partially occupied. 
Due to spin-orbit interaction, the orbital degeneracy in $e_g^\sigma$ is lifted like the $e_g^{\pi}$ case, which allows XMCD arising from $T_z$ (see the inset of Fig.~\ref{fig:xas_xmcd_Cr1}((a)).
As shown in Fig. \ref{fig:xas_xmcd_Cr1}(b), the energy gap between the ground state and the first excited state  ($\Delta E$) increases with stronger spin-orbit interaction.
In Cu$^{2+}$, the expectation value of $T_z$ decreases with increasing spin-orbit interaction, in contrast to Cr$^{2+}$. 
This behavior is presumably due to enhanced mixing with the majority-spin states induced by the spin-orbit interaction, which weakens the overall spin polarization \cite{sasabe2023ferroic}.
Since the first excited initial state possesses an inverted XMCD signal, and $\Delta E$ for $d^4$ and $d^9$ are smaller compared to those for other ions (see Table 1), the XMCD rapidly disappears as the temperature increases. 
Figure 6 shows the XMCD spectra of Cr$^{2+}$ and Mn$^{2+}$ for both the ground state and the first excited state, as well as the results obtained by mixing the initial states according to a Boltzmann distribution with temperature. 
Here, it is assumed that the internal magnetic field of the antiferromagnet does not change with temperature. 
In Cr$^{2+}$ of Fig.~\ref{fig:xas_xmcd_Cr2}(a), the XMCD decreases sharply with increasing temperature, whereas in Mn$^{2+}$ of Fig.~\ref{fig:xas_xmcd_Cr2}(b) the first excited state lies at a higher energy and has a similar spectral shape, making the XMCD more robust against temperature.
As shown in Fig.~\ref{fig:xas_xmcd_Cr1}(b), the energy gap increases with stronger spin-orbit interaction, suggesting that XMCD signals may persist up to higher temperatures in $4d$ and $5d$ transition metal systems.

\section*{Conclusion}
In this study, we have theoretically demonstrated that X-ray magnetic circular dichroism (XMCD) can emerge in collinear altermagnets with trigonal crystal fields, despite the absence of net magnetization. 
The key mechanism originates from the anisotropic magnetic dipole operator $T_z$, which arises from the coupling between spin and quadrupolar orbital distributions. 
This contribution remains finite when the N\'eel vector lies in the basal plane, reflecting the underlying crystal symmetry.
By constructing a complete multipole basis and analyzing the orbital wavefunctions under trigonal distortion, we showed that spin-orbit interaction leads to a nonzero expectation value of $T_z$, thereby enabling a finite XMCD response.
Our calculations, based on both one-electron and multi-electron models incorporating Coulomb interactions, reproduce finite XMCD spectra across a range of $3d^n$ configurations. 
In particular, we find that even ions with $\langle T_z \rangle = 0$ in the ground state (e.g., Mn$^{2+}$, V$^{2+}$) can exhibit XMCD due to lifted degeneracy in the final states excited by the dipole transition. 
These results emphasize that XMCD is a sensitive probe of local electronic anisotropy, even in systems without spin or orbital moments.
Our findings not only clarify the microscopic origin of XMCD in $\alpha$-MnTe, but also predict similar phenomena in other antiferromagnets with the same symmetry, such as FeS. Due to the present choice of model parameters, the XMCD for Ni$^{2+}$ is very weak; nevertheless, we emphasize that strong $T_z$-driven XMCD could also emerge in Ni compounds under trigonal distortions. 
This work broadens the applicability of XMCD beyond ferromagnets and opens up new possibilities for probing hidden magnetic multipoles in exotic antiferromagnetic and altermagnetic systems \cite{SeoJPSJ2021, Naka2021PRB, NakaPRB2022, MisawaPRB2023, RimmlerNatRev2025, Naka2025npj}.

\section{Acknowledgment}
The authors thank T. Uozumi for the productive discussion.
This project is partly supported by the Japan Society for the Promotion of Science (JSPS) KAKENHI (19H04399, 24K03205, 24H01685, and JP25K03387).
This work was supported by MEXT Quantum Leap Flagship Program (MEXT Q-LEAP) Grant Number JPMXS0118068681.
This work is also partially supported by CREST(JPMJCR2435), Japan Science and Technology Agency (JST).

\bibliographystyle{apsrev4-2}
\bibliography{bib_ssb}

\clearpage

\end{document}